# Non-equilibrium Solute Capture in Passivating Oxide Films

Xiao-xiang Yu[1]†, Ahmet Gulec[1]†, Quentin Sherman[1], Katie Lutton Cwalina[2], John R. Scully[2], John H. Perepezko[3], Peter W. Voorhees[1] and Laurence D. Marks[1*]

[1]Department of Materials Science and Engineering, Northwestern University, Evanston, IL 60208, USA.

[2]Department of Materials Science & Engineering, University of Virginia, PO Box 400745, 395 McCormick Road, Charlottesville, VA 22904, USA.

[3]Department of Materials Science and Engineering, University of Wisconsin-Madison, 1509 University Avenue, Madison, WI 53706, USA.

*Correspondence to: L-marks@northwestern.edu.

†These authors contributed equally to this work.

## Abstract

If all humans vanished tomorrow, almost every metal structure would collapse within a century or less, the metal converting to an oxide. In applications ranging from the mature technology of nuts and bolts to high technology batteries, nuclear fuels and turbine engines, protective oxide films are critical to limiting oxidation. To date models of these oxide films have assumed that they form thermodynamic equilibrium stable or metastable phases doped within thermodynamic solubility limits. Here we demonstrate experimentally and theoretically the formation of unusual non-equilibrium oxide phases, that can be predicted using a scientific framework for solute capture at a moving oxide/substrate interface. The theory shows that solute capture is likely a generic process for many electrochemical processes, and suggests that similar phenomena yielding non-equilibrium phases can occur and be predicted for a wide range of other processes involving solid-fluid and solid-solid chemical reactions.

Every metal, with the sole exception of gold, is unstable in air relative to its oxide at room temperature. The only reason they can be used is because of protective oxide films. These protective films are critical, and a strong case has been made by Macdonald[1] that modern civilization would not be the same if they did not exist. Although many of the general concepts of how they form are known, there are still many gaps. Much of our current knowledge is from the mesoscale down to several nanometers, but multiple processes occurring over wide spatial and temporal scales control the nucleation, composition, stability, and structure of the oxide scales which prevent further oxidation of the underlying metal. Understanding the oxide growth at the (sub) nanometer scale is critically important if we are to move beyond simple cost-prohibitive remedies, such as painting surfaces or empirical alloy composition selections, towards more science guided alloy design. This is becoming increasingly urgent with aging infrastructure in the developed world, extension of materials to increasingly harsh environments as well as the ever increasing costs of failure of corroded components such as oil or gas pipelines to name just a few examples.

We demonstrate here experimentally and theoretically that existing models of oxide growth during are incomplete, and in some respects misleading because they do not account for critical kinetic phenomena. We demonstrate that unusual combinations of structure and chemical composition are a general phenomenon and provide a predictive explanation by analogy with the well-established science of solute trapping[33] extended to moving oxide/substrate interfaces. We will call this "Non-equilibrium Solute Capture" to distinguish it, since the differences are large; for instance, solute capture can occur when an interface is physically stationary.

More details will be described below and in the Supplemental Material, we will briefly prequel Non-equilibrium Solute Capture. One has an alloy which is oxidizing with a moving oxidation front that combines point defect migration and physical motion of the interface. The interface therefore has an effective velocity proportional to the rate of incorporation of alloy atoms as cations in the oxide. As the velocity tends to infinity, metal atoms in the alloy cannot achieve equilibrium configurations, they are captured in the oxide with the alloy composition, *independent of what crystallographic phase forms*. In the other limit, as the velocity tends to zero there is a thermodynamic driving force for the alloy atoms to form the stable or metastable phases found in conventional phase diagrams, with compositions bounded by solubility limits. With intermediate velocities, solute atoms are captured in the oxide with compositions far in excess of the solubility limits. The compositions formed can be predicted and calculated using the Non-equilibrium Solute Capture framework based solely upon the ratio of the velocity of the interface and an effective velocity for equilibration.

As general background, the established models of protective film oxide formation in aqueous environments[2] such as salt water as well as during oxidation at higher temperatures have focused upon the development of equilibrium stable or metastable phases present in established phase diagrams, for instance nickel oxide and chromia for the case of nickel-chromium based alloys[3-7] or enrichments of alloying elements in passive films[8-11]. A thin oxide film develops first with a substantial potential gradient across it, as first described by Cabrera and Mott[12]. As the film thickness increases point defects (for instance metal vacancies) diffuse across the oxide due to both the chemical potential gradients and the electric fields[12-17], and the film thickness will eventually exceed a limit where diffusion is now the controlling factor as first described by Wagner[18]. In an aqueous environment the film may never reach this limit due to simultaneous dissolution at the oxide/water interface[19,20]. In some cases a thin oxide film develops which does not significantly oxidize further leading to a self-protective coating.

Throughout the existing literature, thermodynamically favored oxide phases are commonly assumed to form (e.g.[21-24]). The conventional view is that during oxidation the more energetically favored (exothermic) oxides form before those which are less exothermic (e.g.[25]). These oxides may be doped, for instance some chromium in the nickel oxide but only within equilibrium solubility limits. For aqueous films there is extensive x-ray photoelectron data (e.g.[21,26]) which has been interpreted in terms of the thermodynamically stable oxides; for oxidation x-ray or electron diffraction (e.g.[27-30]) is typically used to identify the lattice parameters of the phases which are then compared to the known oxides in the phase diagrams. While some more recent in-situ work has indicated that the most energetically favored oxide often do not form first (e.g.[31]), the focus has been on thermodynamically stable or metastable phases of equilibrium compositions that appear in conventional phase diagrams of the oxides.

Is this model complete? Across many areas of materials science it is well established that kinetics matters as much if not more than thermodynamics. Kinetic control is ubiquitous, ranging from the industrial production of chemicals where the presence of a catalyst leads to selective formation of some desired chemical product rather than the thermodynamic endpoint, to *in-vivo* protein mediated biochemistry or the formation of many if not most technological materials such as complex multilayer electronic devices. For instance, for thin film deposition it is known that kinetics can lead to phases far from equilibrium that cannot be produced by other means. However, materials far from equilibrium are commonly considered to be the exception.

Key to understanding the formation of the protective oxides is knowledge of the phases present, for which *both* structural and chemical information are required; each alone can be misleading. We are aware of only one paper[32] which has looked carefully at this for oxidation, finding unexpected compositions that are not present in any of the thermodynamic databases for the relevant oxides. The authors of this paper[32] were not able to explain the source of their apparently anomalous result, and in the forty years since publication these results seem to have been ignored by the larger scientific community.

The systems we will focus upon here are Ni-Cr-Mo alloys oxidized in air at intermediate temperatures in the range 500-700 ºC, as well as the same materials passivated electrochemically in aqueous environments; further details can be found in the Methods section. While specific details vary with the particular alloys and treatment conditions, we will demonstrate that solute capture during oxide film growth is a general process which we believe matters for many, if not all, other metal alloys and many other electrochemical processes. The main experimental tools used are transmission electron microscopy and atom-probe tomography. We have examined both bulk samples which were oxidized and then appropriate regions extracted using conventional sample preparation methods, as well as samples that were first prepared in the appropriate geometry for examination and then oxidized or corroded. The latter are important as they allow us to exclude artifacts during the sample preparation and the evolution of stresses.

Independent of whether the oxide is formed by dry oxidation or in aqueous conditions, the first product formed is an oxide with a rocksalt structure ($Fm\overline{3}m$) which is normally interpreted in the literature to be nickel oxide since it has essentially the same lattice parameter, with perhaps some additional metal atoms as low concentration dopants. From electron diffraction data as well as high resolution imaging (see fig. S1) the crystallographic structure is rocksalt, but a closer examination using both atom probe tomography and chemical analysis inside electron microscopes (see Fig. 1 for passive film formed electrochemically) shows that there is very significant chromium and traces of molybdenum in the rocksalt oxide. This indicates that the chromium and molybdenum are located as substitutional atoms. Ignoring for simplicity the low molybdenum concentration, from the experimental data the composition of this rocksalt structure is $Ni_{1-x}Cr_xO_y$ with x~0.5. (For the thin regions the oxygen content is hard to quantify; it tends towards y=1.5 at the outer surface and y=1 adjacent to the metal or for very thin oxides.) The chromium content is significantly larger than the literature equilibrium thermodynamic phases of NiO, $Cr_2O_3$ and the spinel $NiCr_2O_4$ for which the literature indicates only small solid solution ranges[34].

We should stress that our results indicate that one can have a rocksalt phase ($Fm\overline{3}m$) with a NiO composition in some cases, for instance in Fig. 3 which will be discussed below; the key science is the detection of the unexpected $Ni_{1-x}Cr_xO_y$ composition with the same space group.

The rocksalt phase is not the only structure we have found experimentally with a composition far from conventional thermodynamic expectations. For nickel-chromium-iron alloys and nickel-chromium-molybdenum-alloys an oxide with the corundum structure ($R\overline{3}c$) has been reported[21,26,35-38], and assumed to be chromia. We can verify the existence of an oxide with a corundum structure consistent with the existing literature, in some cases it is chromia $Cr_2O_3$ but in other cases careful compositional analysis demonstrates a composition of $Ni_xCr_{2-x}O_{3-y}$ with x~1 as shown in Fig. 2, here for a sample oxidized in-situ at 700 ºC in 1x10$^{-4}$ torr of oxygen. A metastable $R\bar{3}c$ Ni$_2$O$_3$ phase with a corundum structure is well established[39], so a metastable corundum Ni$_x$Cr$_{2-x}$O$_{3-y}$ is structurally and chemically reasonable, but far from expectations based upon the published thermodynamically stable phases.

The examples in Figs. 1 and 2 are for relatively thin films, so it could be that these unexpected compositions are just transients that are not relevant for thicker films. This is not the case. Shown in Fig. 3 are results for a sample that was annealed in oxygen for 24 h at 800°C, where similar results are found for oxide films more than 100 nm thick. Adjacent to the metal is an approximately 100 nm thick Ni$_{1-x}$Cr$_x$O$_{1+x/2}$ rocksalt structure (with solute captured Cr), then over this is nearly pure rocksalt NiO with Cr$_2$O$_3$ at the outer surface. The evidence indicates that given enough time the metastable Ni$_{1-x}$Cr$_x$O$_{1+x/2}$ rocksalt phase away from the metal/oxide interface can transform to NiO and Cr$_2$O$_3$, but the rocksalt oxide near the metal/oxide interface still has a non-equilibrium composition.

While the most definitive current evidence we have is from local structural and chemical analysis, in-situ atomic emission spectro-electrochemistry experiments support this conclusion as well. These involve tracking the metal ions leaving the oxide at the oxide-solution interface versus those oxidized[40] and are described in more detail in the Supplemental Material §S2 and Supplemental figs S2-S5. If we use a conventional interpretation where the oxides being formed are rocksalt NiO and corundum Cr$_2$O$_3$, the experimental data does not balance mass. If instead we analyze the data for all the Ni and Cr in the oxide based on a combination of the two phases detected above, namely rocksalt Ni$_{1-x}$Cr$_x$O with x=0.2 and Ni$_x$Cr$_{2-x}$O$_{3-y}$ with x=1 and y=0 we obtain semi-quantitatively a valid mass balance for Cr, Ni and O.

All the experimental evidence points towards non-equilibrium phases consisting of both the rocksalt and corundum phases with (based upon the metal/oxygen ratios) either Ni$^{2+}$ or Ni$^{3+}$ with dominantly Cr$^{3+}$ but a possibility of Cr$^{2+}$ very close to the metal/oxide interface. Note that the rocksalt and corundum phases are not that different, rocksalt has an ABC cubic packing of hexagonal planes, whereas corundum has an AB hexagonal packing. In both the metal occupy octahedral sites and the metal-oxygen distances are very similar for Ni$^{2+}$ and Cr$^{3+}$; small changes are needed to accommodate any changes in co-ordination. Density functional calculations (see Supplemental Material) indicate that the preferred spin configuration is antiferromagnetic.

These highly non-equilibrium phase compositions, well beyond established solubility limits, are a result of what we will call non-equilibrium solute capture at moving oxidation fronts, analogous but different from solute trapping[33]. Differentiating the two is important. In classical solute trapping atoms add to a moving solidification front, so the velocity of the front is directly coupled to the rate of addition of atoms. In contrast, in non-equilibrium solute capture in addition to a moving interfaces both vacancies and interstitials are moving so the fluxes of chemical species and the velocity of the interface are decoupled[41]. (More details are provided in Supplemental Material §S3.) The important term is the combination of the large fluxes of atoms at the oxide-metal interface and motion of the interface compared to the diffusion rate of solute atoms. An important measure of the extent of capture in an effective medium model is $\beta = v^{Eff}/v^{Eq}$, where $v^{Eff}$ is the effective velocity of the interface combining both physical motion and the flux of atoms, and $v^{Eq}$ is an effective velocity for equilibration. The later will be the sum of a two terms, $v^{Eq} = v^{Exch} + v^{Nuc}$. The first, $v^{Exch}$, is similar to that in solute trapping $v^{Exch} = a/D_i$ where $D_i$ is an interface diffusion coefficient for a Zener exchange of the relevant atoms and a is the hopping distance for exchange of atoms between the oxide and metal to equilibrate the composition of the oxide. The second will be an effective velocity for nucleation of phase separation, for instance rocksalt $Ni_{1-x}Cr_{2x/3}O$ phase separating into rocksalt NiO and corundum $Cr_2O_3$ doped within solubility limits. We argue that similar to classic solute trapping, when $\beta$ is significantly larger than unity the equilibrium interfacial concentrations do not develop; when it is significantly smaller than unity local thermodynamic equilibrium can hold at the interface in which case the compositions are well described by those on the phase diagram. Non-equilibrium phase compositions can form when $\beta \gg 1$ so long as the total free energy change of the phase transformation is negative[33].

We argue that the combination of an oxide front moving into the metal and the large fluxes crossing the interface for both oxidation and in aqueous environments leads to capture in the oxide of compositions at much higher concentrations than equilibrium thermodynamics predict as illustrated in Fig. 4. The thermodynamic requirements for solute capture in the specific case of Ni-Cr oxidation are discussed in more detail in Supplemental Material §S3, where it is shown that almost complete capture in either rocksalt or corundum structures is possible for a wide range of conditions of temperature and pressure during oxidation and during aqueous corrosion and passivation.

Turning to the values of $\beta$, for planar growth of NiO and using standard values for diffusion constants in the oxide (e.g.[42,43]). Fig, 5A shows that the capture of Cr in NiO is to be expected for a wide range of conditions (see also Supplemental Material §S4). It is also probable that $\beta>1$, during electrochemical passivation as shown in Fig. 5B. In every case except very slow passive aqueous dissolution at $<10^{-7}$ A/cm$^2$, the effective velocity of the corroding interface given in 5B using a Faraday's law derived electrochemical penetration rate is faster than the rate of equilibration.

The model for solute capture in oxides has an important consequence; it predicts that for most cases of oxidation and almost all aqueous conditions, solute capture is the norm, not the exception. This result has significant consequences for our understanding of protective oxide films, as well as opening the door to new routes to improving the resistance of materials for applications ranging from hip-implants to batteries, turbine blades or lower technology uses such as cookware. We will discuss some of the science first. Rather than having a simple, insulating oxide, one has a complex highly doped semiconductor which could even reach the doping levels of a degenerate semiconductor (metal). As an illustration of this, fig S7 shows the calculated density-of-states for solute trapped Cr and Mo in rocksalt NiO; in both cases additional states are introduced in the gap, which will change the band-bending in the material. Extra states will significantly change the electric field across the oxide, and the fundamental semiconductor physics of the protective oxide film. These need to be included into the classic approach of Cabrera and Mott[12] or other variations (e.g.[13-17,44]).

In addition to electronic changes one also has to consider how the film develops and changes as a function of time. For instance, in aqueous passive film breakdown, metastable pitting and recovery by repassivation are typically considered to be processes involving very rapid oxide growth, whereas quasi steady-state passivation film growth is slow[45]. During slow growth the film structure has the opportunity to "age", similar to aging of metal alloys where time is provided for the creation of additional phases, here phase separation of the solute captured atoms. However, since passive films are in a constant state of breakdown and repair as well as dissolution and oxidation at the film/solution and metal/film interface to just maintain a constant thickness, the solute captured condition is almost always the most relevant.

There is also much more to be done with both the experimental and theoretical analysis of solute capture for moving oxide fronts and other electrochemical applications. For instance, it is clear from the atom probe data in Fig. 4F that there is partial clustering of the metal atoms, it is not a pure random solid solution. The experimental data in the line scans also indicates some chemical gradients of the metal atom concentrations within the oxides, which whilst not unexpected merits much more exploration.

The new view of these protective oxide films opens the door to new strategies to improve alloys to yield higher performing protective oxide films. It is important to note that the extent of capture can be predicted. The concentration of the growing oxide depends on the free energies of the phases, which are sometimes known or can be estimated well with advanced density functional methods (see Supplemental Material §S4), as well as values for the interface diffusion constant and hopping distance. While the later are not yet known for solute capture, they can be calculated from atomistic methods. Thermodynamic databases along with the model for solute capture during oxidation will allow oxides of novel compositions, and thus properties, to be engineered for a given application. We believe it should be possible to design solute captured protective oxides so, for instance, they better resist breakdown phenomena or provide better protection for high-temperature turbine engines to have significantly longer lifetimes in service.

Finally, in terms of broader materials science we strongly suspect that comparable solute capture can occur for many other solid-fluid or solid-solid chemical reactions, not just the formation of protective oxide films or the classic case of rapid solidification. If the relevant product is lower in free energy, and the reaction front is moving fast enough, the same fundamental materials science should hold independent of whether one is dealing with ceramics, metals, semiconductors or polymers.

**Figures with Figure Captions**

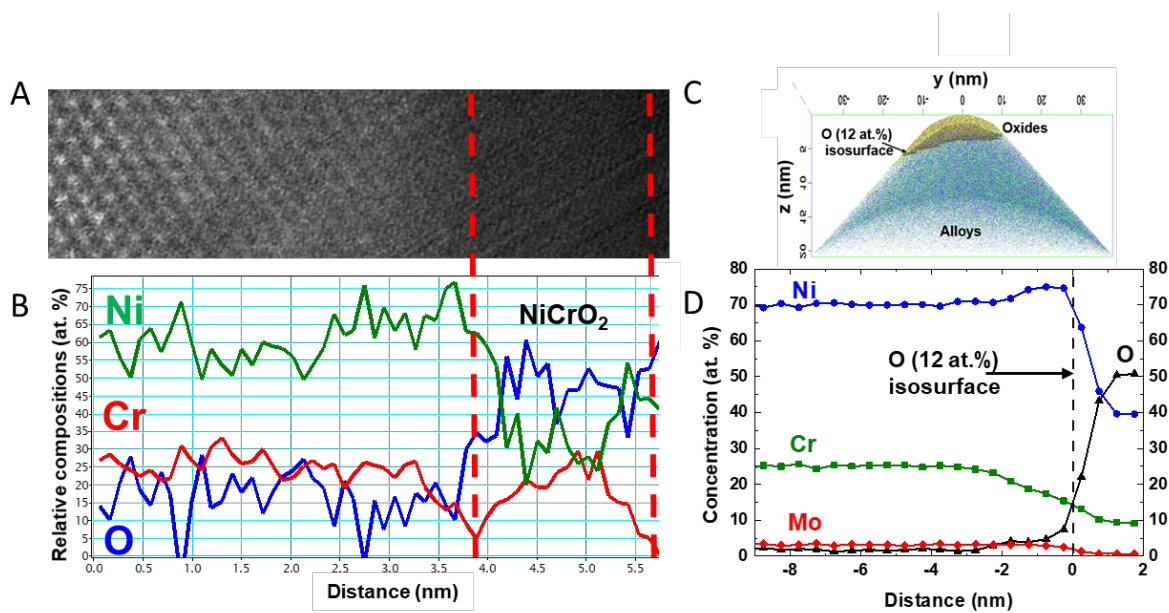

**Figure 1** Chemical data for the oxide formed on a NiCrMo alloy oxidized in $K_2S_2O_8$ and $Na_2SO_4$. In (A) a high-angle annular dark field image is shown, with (B) the composition measured from an electron energy loss line-scan. The oxide immediately on the metal has a composition of approximately $Ni_{0.5}Cr_{0.5}O$. Shown in (C) and (D) are atom-probe tomography results of a tip treated under the same conditions with a 12 at % isosurface in c) and a proxigram (compositional line scan) in (D), both corrected for the oxygen detection efficiency. While the structural data is not present in the atom-probe data, the compositional information is and cross-validates the electron microscopy results.

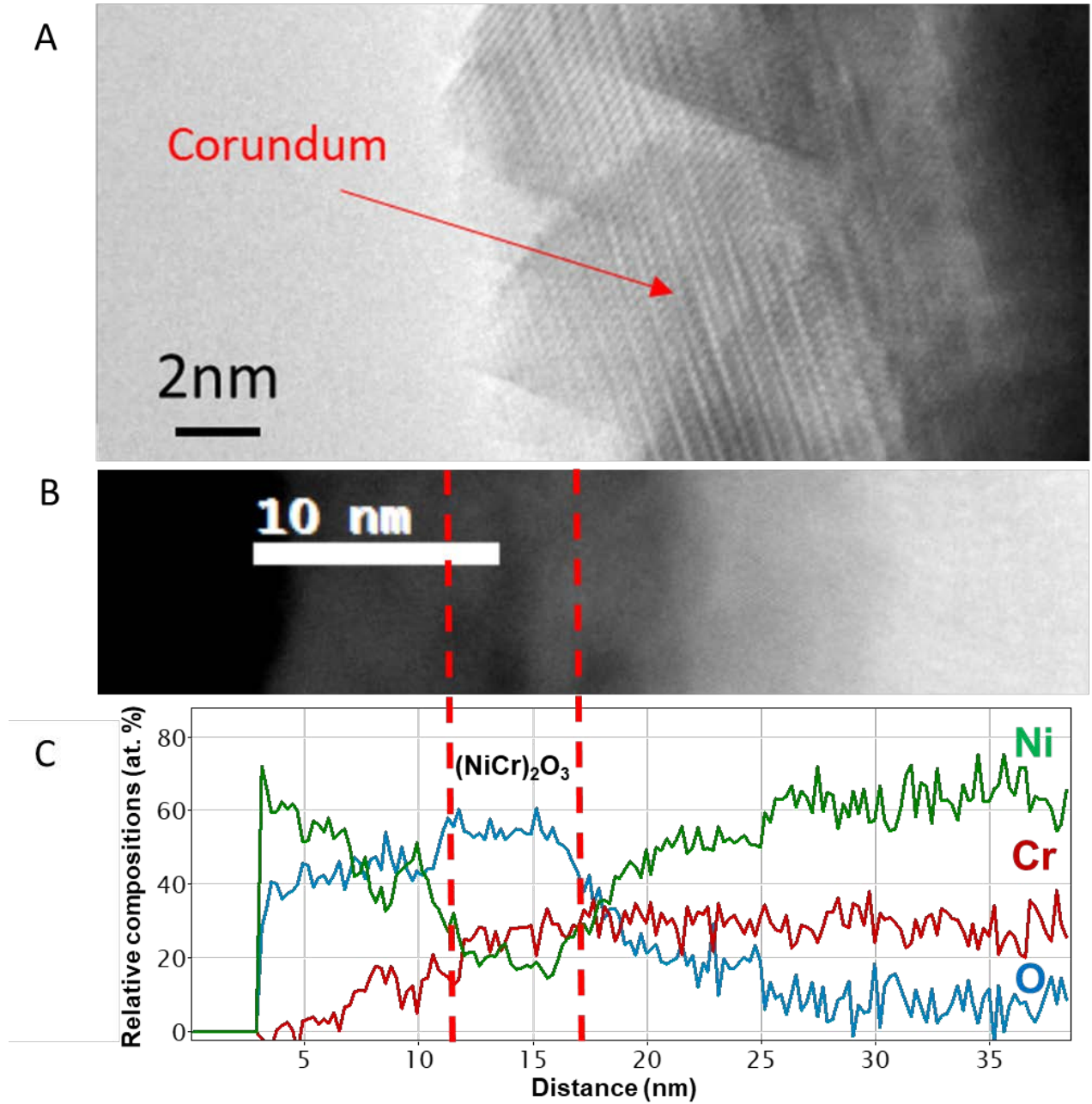

**Figure 2** Results for a sample oxidized in-situ at 700 C in 1x10$^{-4}$ torr of oxygen, and later analyzed by electron energy loss spectroscopy. In (A) is shown a bright field image during in-situ growth where fringes characteristic of the corundum structure are apparent. Shown in (B) is a high-angle annular dark field image with the corresponding electron energy loss line scan in (C). The corundum structure adjacent to the metal has a composition of approximately NiCrO$_3$.

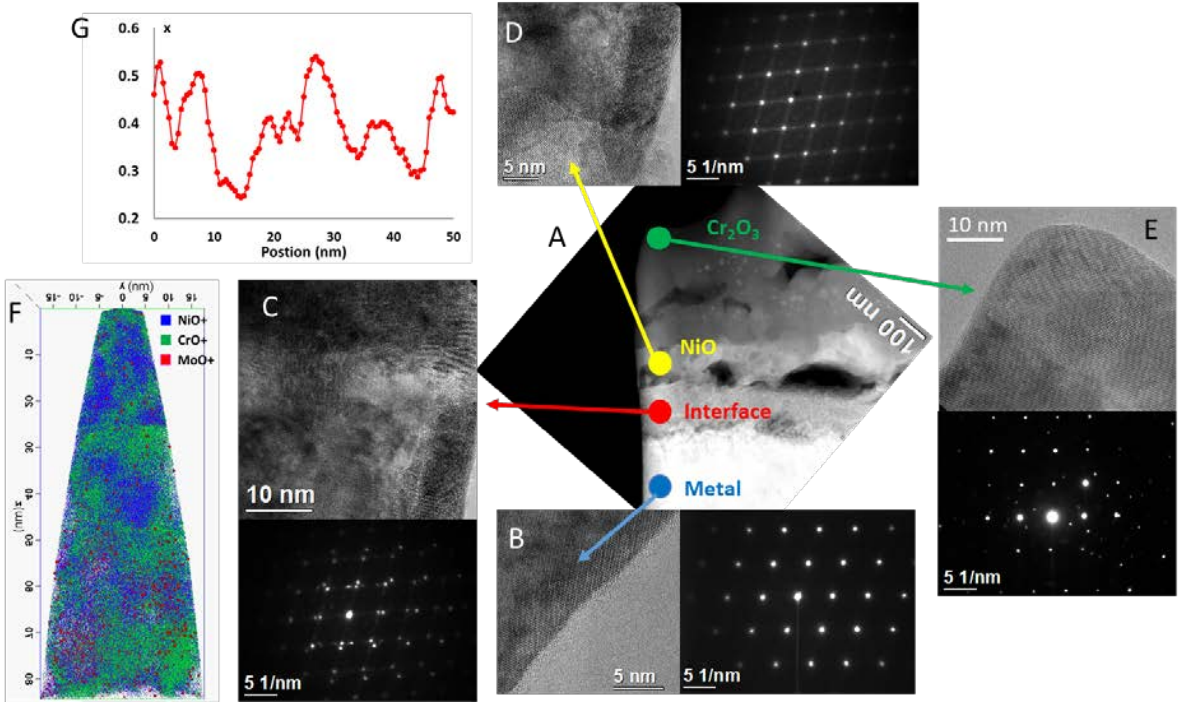

**Figure 3** Montage of results for a sample oxidized at 800 C for 24 h in air. Shown in (A) is an overview high-angle annular dark field image of a focused ion-beam region cut from the sample. The different layers in (B)-(E) are shown as bright-field and diffraction pattern pairs, in (B) of the metal, (C) the rocksalt structure, (D) nickel oxide and (E) chromia. Shown also in (F) is an atom-probe tomogram taken from the interface layer which shows, similar to Fig 1, the presence of a $Ni_{1-x}Cr_xO_{1+x/2}$ rocksalt phase, with a one dimensional composition plot with the metal on the right in (G) which shows metal composition variations (x) with $Ni_{1-x}Cr_xO_{1+x/2}$ composition.

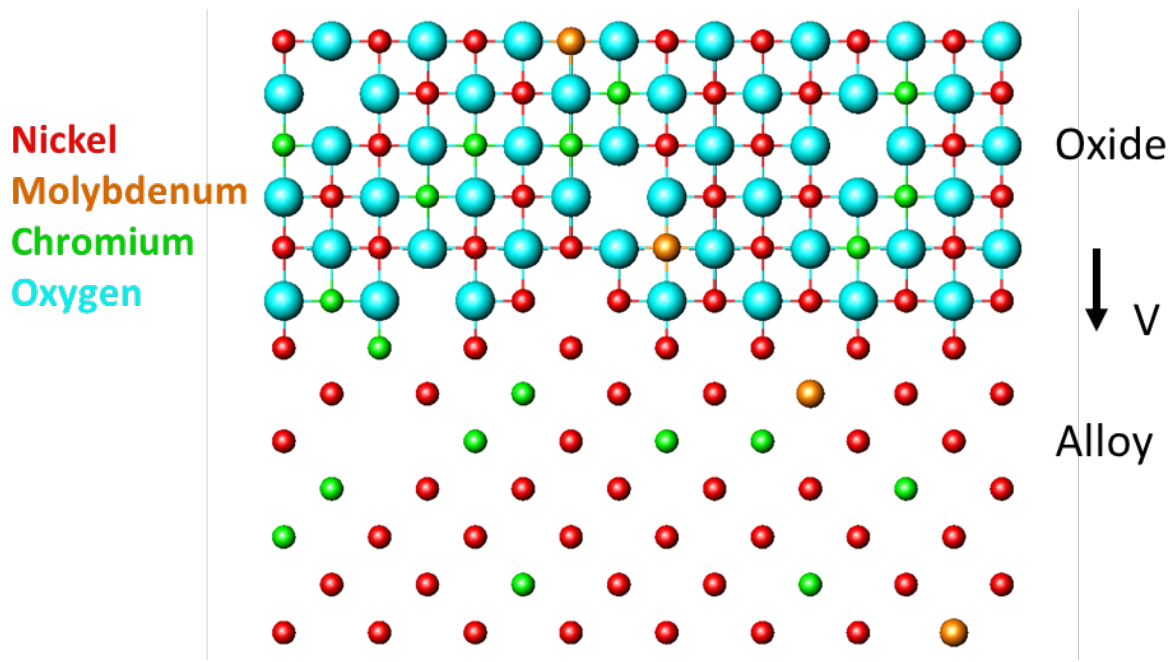

**Figure 4** Representation of an oxidation front with an oxide (top) moving into an alloy (bottom) with a velocity v, with the different types of atoms color coded, with a number of metal vacancies. Note that the interface velocity and vacancy flux will in general be different.

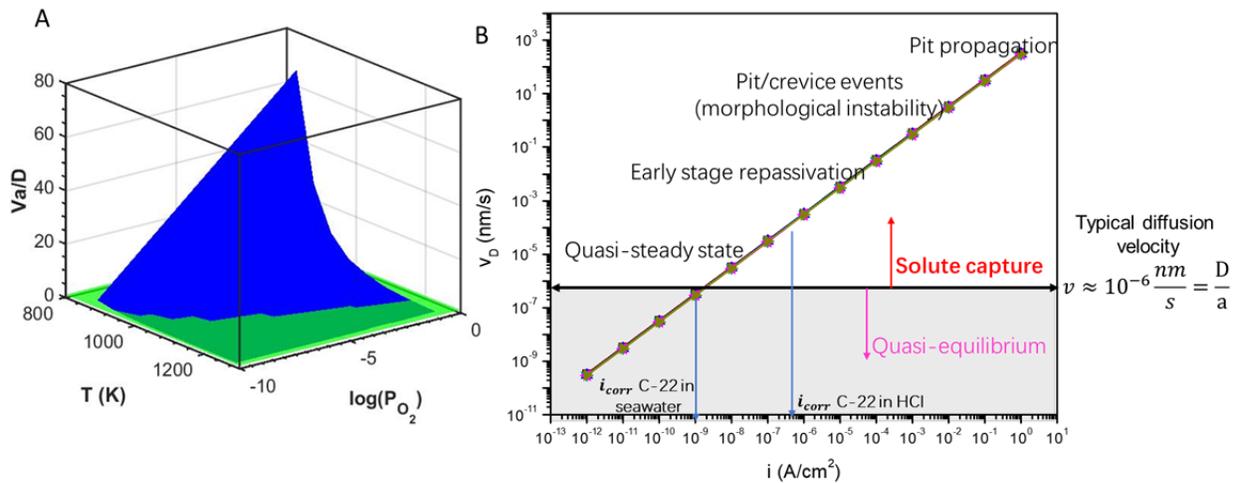

**Figure 5** Plots of the kinetic conditions for solute capture, in (A) for oxidation using kinetics based upon literature diffusion constants and reaction rates as a function of exposure temperature and oxygen partial pressure in atmospheres. The plot is for an oxide thickness of 5nm and a hopping distance of 1nm. The horizontal green plane is when $v^{eff}a/D_i=1$, the blue surface the ratio for different temperatures and oxygen pressures. For a wide range of conditions, high temperature oxidation is expected to lead to solute capture, consistent with the experimental results. In (B) a plot of the local electrochemical corrosion current density along x and the local velocity of the oxidation front along y for a range of different processes in aqueous corrosion with the approximate diffusion velocity $D_i/a$ marked on the right for a=1.0 nm and assumed a field affected diffusivity of $10^{-20}$ cm$^2$/sec, showing that solute capture will be common for many conditions.

## METHODS

**Electrochemical treatment:** Ni-Cr and Ni-Cr-Mo alloys were oxidized (passivated) electrochemically using a potentiostat in 0.1 M NaCl solution adjusted with 0.1 mM HCl to pH 4, Samples with any air-formed oxides where were reduced for 10 min at -1.3 V vs. SCE ( Saturated Calomel) for 10 minutes and then polarized potentiostatically to +0.2 $V_{SCE}$. TEM and 3DAPT specimens where prepared for each method and to avoid artifacts were then oxidized over the entire exposed surface as a last step using a chemical potentiostat. 200 ml of a 0.1 M NaCl solution adjusted with 0.1 mM HCl to pH 4 was treated with 1-5 μL of $H_2O_2$ (30%) as an oxidizing agent or in 0.1 M $Na_2SO_4$ with a few uL of $K_2S_2O_8$. This condition was verified to produce an electrochemical potential near +0.2 $V_{SCE}$. Verification of nearly identical oxides from this method and the potentiostat approach was achieved by electrochemical impedance. In both cases specimen were oxidized for 10,000 seconds or more.

**Transmission electron microscopy**: TEM samples were prepared using a low speed saw (Buehler, Isomet) to cut the bulk sample into thin slices (~0.7 mm), then mechanical thinned to less than 100 um using sandpapers. From the thin slices, several 3 mm disks were cut with a disk cutter (South Bay Technology, Model 360) and dimpled (VCR group, Inc., D500i) to the thickness of 10~20 um. Finally, the samples were ion milled at 3~6 keV (Gatan precision ion polishing system, Model 691) until a small hole appeared, followed by low energy and low angle argon ion milling (Fischione Model 1040 Nano-Mill) at 500 eV to remove the amorphous and implanted layers. After preparation the samples were either electrochemically treated (above) or oxidized.

Atomic resolution scanning transmission electron microscope imaging was performed using an aberration corrected JEOL ARM200CF with collections angles $\beta$ of $11 mrad \leq \beta \leq 22 mrad$ for annular bright field (ABF) imaging and $90 mrad \leq \beta \leq 220 mrad$ for high angle annular dark field (HAADF) imaging, these being acquired simultaneously. Compositonal analysis was performed with a Gatan Efina EELS and an Oxford X-max 100TLE windowless SDD X-ray detector attached to the ARM200CF. Electron diffraction and high-resolution imaging experiments were performed utilizing both a JEOL ARM200CF and JEOL 2100F.

**Atom Probe Tomography:** The samples for atom probe tomography were prepared by a standard electropolishing procedures: 0.5mm×0.5mm×2cm alloys bars were electropolished in an electrolyte of 10% (vol %) perchloric acid in acetic acid using 12V~20V DC for pre-thinning. Then 2% (vol %) perchloric acid in a butoxyenthanol electrolyte and 12V~15V DC were used for the final thinning. A CAMECA Local Electrode Atom Probe (LEAP) 4000XSi with an ultrafast detector capability was used for APT experiments. Picosecond pulses of ultraviolet laser light were utilized to evaporate individual atoms at a pulse repetition rate of 250 kHz, a laser pulse energy of 20 pJ per pulse and an average detection rate of 0.005 ions per pulse. The specimen tip temperature was maintained at 25K. Data analyses were performed on the three-dimensional (3-D) reconstructions of specimens utilizing the program IVAS 3.6.1.

It is established that the composition for oxides measured using Atom Probe depends upon the laser energy per pulse, tending to accurate as this is reduced. To calibrate the data, a tip of pure nickel was prepared and then oxidized and used as a calibrant. This sample indicated that the detection efficiency of oxygen was 2/3. All the data shown in Figures has been corrected for the detection efficiency of the oxygen.

## References

[1] Macdonald, D.D., Passivity - the key to our metals-based civilization. *Pure and Applied Chemistry* 71 (6), 951-978 (1999).
[2] Beverskog, B. & Puigdomenech, I., Pourbaix diagrams for the ternary system of iron-chromium-nickel. *Corrosion* 55 (11), 1077-1087 (1999).

3. Croll, J.E. & Wallwork, G.R., High-Temperature Oxidation of Iron-Chromium-Nickel Alloys Containing 330percent Chromium. *Oxid Met* 4 (3), 121-140 (1972).
4. Dalvi, A.D. & Coates, D.E., A review of the diffusion path concept and its application to the high-temperature oxidation of binary alloys. *Oxid Met* 5 (2), 113-135 (1972).
5. Gilman, J.J., Oxide surface films on metal crystals. *Science* 306 (5699), 1134-1135; author reply 1134-1135 (2004).
6. Ziemniak, S.E., Anovitz, L.M., Castelli, R.A., & Porter, W.D., Thermodynamics of $Cr_2O_3$, $FeCr_2O_4$, $ZnCr_2O_4$, and $CoCr_2O_4$. *J Chem Thermodyn* 39 (11), 1474-1492 (2007).
7. Kaufman, L. *et al.*, Transformation, stability and Pourbaix diagrams of high performance corrosion resistant (HPCRM) alloys. *Calphad* 33 (1), 89-99 (2009).
8. Kirchheim, R. *et al.*, The Passivity of Iron-Chromium Alloys. *Corrosion Science* 29 (7), 899-917 (1989).
9. Zhang, L. & Macdonald, D.D., Segregation of alloying elements in passive systems - I. XPS studies on the Ni-W system. *Electrochimica Acta* 43 (18), 2661-2671 (1998).
10. Castle, J.E. & Asami, K., A more general method for ranking the enrichment of alloying elements in passivation films. *Surf Interface Anal* 36 (3), 220-224 (2004).
11. Macdonald, D.D., The role of passivity in the corrosion resistance of metals and alloys. *Afinidad* 62 (519), 498-504 (2005).
12. Cabrera, N. & Mott, N.F., Theory of the Oxidation of Metals. *Reports on Progress in Physics* 12, 163-184 (1948).
13. Chao, C.Y., Lin, L.F., & Macdonald, D.D., A Point-Defect Model for Anodic Passive Films .1. Film Growth-Kinetics. *Journal of the Electrochemical Society* 128 (6), 1187-1194 (1981).
14. Lin, L.F., Chao, C.Y., & Macdonald, D.D., A Point-Defect Model for Anodic Passive Films .2. Chemical Breakdown and Pit Initiation. *Journal of the Electrochemical Society* 128 (6), 1194-1198 (1981).
15. Atkinson, A., Transport Processes during the Growth of Oxide-Films at Elevated-Temperature. *Rev Mod Phys* 57 (2), 437-470 (1985).
16. Macdonald, D.D., The Point-Defect Model for the Passive State. *Journal of the Electrochemical Society* 139 (12), 3434-3449 (1992).
17. Macdonald, D.D., The history of the Point Defect Model for the passive state: A brief review of film growth aspects. *Electrochimica Acta* 56 (4), 1761-1772 (2011).
18. Wagner, C., The theory of the warm-up process. *Zeitschrift Fur Physikalische Chemie-Abteilung B-Chemie Der Elementarprozesse Aufbau Der Materie* 21 (1/2), 25-41 (1933).
19. Olefjord, I. & Elfstrom, B.O., The Composition of the Surface during Passivation of Stainless-Steels. *Corrosion* 38 (1), 46-52 (1982).
20. Olefjord, I., Brox, B., & Jelvestam, U., Surface Composition of Stainless-Steels during Anodic-Dissolution and Passivation. *Journal of the Electrochemical Society* 132 (12), 2854-2861 (1984).
21. Winograd, N., Baitinger, W.E., Amy, J.W., & Munarin, J.A., X-ray Photoelectron Spectroscopic Studies of Interactions in Multicomponent Metal and Metal Oxide Thin Films. *Science* 184 (4136), 565-567 (1974).
22. Willenbruch, R.D., Clayton, C.R., Oversluizen, M., Kim, D., & Lu, Y., An Xps and Electrochemical Study of the Influence of Molybdenum and Nitrogen on the Passivity of Austenitic Stainless-Steel. *Corrosion Science* 31, 179-190 (1990).
23. Saltykov, P., Fabrichnaya, O., Golczewski, J., & Aldinger, F., Thermodynamic modeling of oxidation of Al-Cr-Ni alloys. *Journal of Alloys and Compounds* 381 (1-2), 99-113 (2004).
24. Ingham, B., Hendy, S.C., Laycock, N.J., & Ryan, M.P., Estimates of thermodynamic stability of iron-chromium spinels in aqueous solution based on first-principles calculations. *Electrochemical and Solid State Letters* 10 (10), C57-C59 (2007).
25. Kjellqvist, L., Selleby, M., & Sundman, B., Thermodynamic modelling of the Cr-Fe-Ni-O system. *Calphad* 32 (3), 577-592 (2008).
26. Machet, A. *et al.*, XPS and STM study of the growth and structure of passive films in high temperature water on a nickel-base alloy. *Electrochimica Acta* 49 (22-23), 3957-3964 (2004).
27. Doychak, J. & Ruhle, M., Tem Studies of Oxidized Nial and Ni3al Cross-Sections. *Oxid Met* 31 (5-6), 431-452 (1989).
28. Doychak, J., Smialek, J.L., & Mitchell, T.E., Transient Oxidation of Single-Crystal Beta-Nial. *Metallurgical Transactions a-Physical Metallurgy and Materials Science* 20 (3), 499-518 (1989).
29. Brumm, M.W. & Grabke, H.J., The Oxidation Behavior of Nial .1. Phase-Transformations in the Alumina Scale during Oxidation of Nial and Nial-Cr Alloys. *Corrosion Science* 33 (11), 1677-1690 (1992).

30. Toney, M.F., Davenport, A.J., Oblonsky, L.J., Ryan, M.P., & Vitus, C.M., Atomic structure of the passive oxide film formed on iron. *Physical Review Letters* 79 (21), 4282-4285 (1997).
31. Luo, L.L. *et al.*, In-situ transmission electron microscopy study of surface oxidation for Ni-10Cr and Ni-20Cr alloys. *Scripta Materialia* 114, 129-132 (2016).
32. Takei, A. & Nii, K., The Growth Process of Oxide Layers in the Initial Stage of Oxidation of 80Ni-20Cr Alloy. *Transactions of the Japan Institute of Metals* 17 (4), 211-219 (1976).
33. Baker, J.C. & Cahn, J.W., Solute Trapping by Rapid Solidification. *Acta Metall Mater* 17 (5), 575-578 (1969).
34. Chen, C.H., Notis, M.R., & Williams, D.B., Precipitation and Solid Solubility in the System Ni0-Cr2O3. *Journal of the American Ceramic Society* 66 (8), 566-571 (1983).
35. Marcus, P. & Grimal, J.M., The Anodic-Dissolution and Passivation of Ni-Cr-Fe Alloys Studied by Esca. *Corrosion Science* 33 (5), 805-814 (1992).
36. Ziemniak, S.E. & Hanson, M., Corrosion behavior of NiCrMo Alloy 625 in high temperature, hydrogenated water. *Corrosion Science* 45 (7), 1595-1618 (2003).
37. Ziemniak, S.E. & Hanson, M., Corrosion behavior of NiCrFe Alloy 600 in high temperature, hydrogenated water. *Corrosion Science* 48 (2), 498-521 (2006).
38. Maslar, J.E., Hurst, W.S., Bowers, W.J., Hendricks, J.H., & Windsor, E.S., Alloy 600 Aqueous Corrosion at Elevated Temperatures and Pressures: An In Situ Raman Spectroscopic Investigation. *Journal of the Electrochemical Society* 156 (3), C103-C113 (2009).
39. Pies, W. & Weiss, A., b1468, II.1.1 Simple oxides: Datasheet from Landolt-Börnstein - Group III Condensed Matter · Volume 7B1: "Key Element: O. Part 1" in SpringerMaterials (http://dx.doi.org/10.1007/10201470_31) (Springer-Verlag Berlin Heidelberg, 1975).
40. Lutton, K., Gusieva, K., Ott, N., Birbilis, N., & Scully, J.R., Understanding multi-element alloy passivation in acidic solutions using operando methods. *Electrochem Commun* 80, 44-47 (2017).
41. Gurtin, M.E. & Voorhees, P.W., The thermodynamics of evolving interfaces far from equilibrium. *Acta Materialia* 44 (1), 235-247 (1996).
42. Chen, W.K., Peterson, N.L., & Robinson, L.C., Chromium Tracer Diffusion in NiO Crystals. *Journal of Physics and Chemistry of Solids* 34 (4), 705-709 (1973).
43. Douglass, D., The oxidation mechanism of dilute Ni-Cr alloys. *Corrosion Science* 8 (9), 665-678 (1968).
44. Seyeux, A., Maurice, V., & Marcus, P., Oxide Film Growth Kinetics on Metals and Alloys I. Physical Model. *Journal of the Electrochemical Society* 160 (6), C189-C196 (2013).
45. Frankel, G.S., Stockert, L., Hunkeler, F., & Boehni, H., Metastable Pitting of Stainless-Steel. *Corrosion* 43 (7), 429-436 (1987).

**Acknowledgments** The authors acknowledge support from ONR MURI "Understanding Atomic Scale Structure in Four Dimensions to Design and Control Corrosion Resistant Alloys" on Grant Number N00014-16-1-2280.

**Author Contributions** XXY and AG performed the oxidation and aqueous treatments of the TEM samples, supervised by LDM, with input from KL and JRS. XXY performed the APT experiments, supervised by LDM. QS performed the solute trapping thermodynamics analysis supervised by PWV with discussions with JHP. KL performed the electrochemical tests supervised by JRS. LDM performed the DFT calculations. All authors contributed to writing of the manuscript.

**Competing Financial Interests:** The authors declare no competing financial interests.

**Author Information** Reprints and permissions information is available at www.nature.com/reprints. The authors declare no competing financial interests. Correspondence and requests for materials should be addressed to L.D. Marks (L-marks@northwestern.edu).

# Supplemental Material: Non-equilibrium Solute Capture in Passivating Oxide Films

X.X. Yu[1]†, A Gulec[1]†, Q. Sherman[1], K. Lutton[2], J.R. Scully[2], J.H. Perepezko[3], P. W. Voorhees[1] and L. D. Marks[1*]

## 1. Supplemental Images of Rocksalt Phase

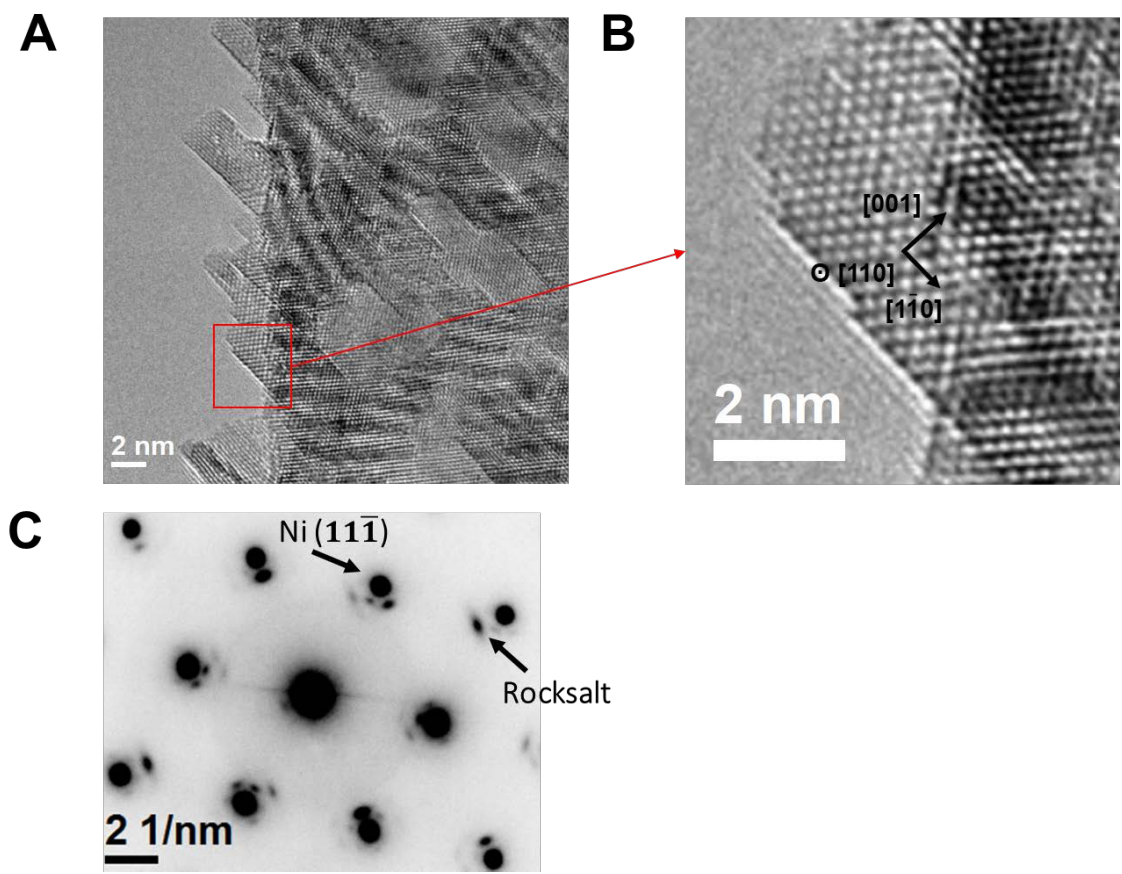

**Fig. S1.** HREM images (A) and (B) show the rocksalt islands grown on Ni-Cr-Mo substrate oxidized in sodium sulfate for 10000 seconds and the diffraction patterns (C) indicate the cube-cube epitaxy growth.

## 2. Electrochemical Tracking Evidence

The kinetics of partial elemental dissolution reactions during electrochemical passivation at +0.2 V vs SCE were analyzed by atomic emission spectroelectrochemistry (AESEC)[1,2]. Previously, AESEC has characterized the *in situ* anodic dissolution of 304 stainless steel, passivation of binary Fe-Cr alloys, and breakdown of passive films on Fe-Cr-Mo glassy alloys[1-3]. Many past studies do not consider the time-dependent contributions of individual elements towards passive film growth and dissolution, as well as the real-time oxide thickness. Measurements of the fate of each alloying element by *operando* electrochemical passivation and film dissolution reactions occurring on Ni-Cr and Ni-Cr-Mo were accomplished using an Inductively Coupled Plasma-Mass Spectrometer (ICP-MS) in a flow cell arrangement[4]. By coupling the ICP-MS measurements to the DC electrochemical measurement of total oxidation current density, $i_{EC}$, and element-specific dissolution reactions could be directly monitored and considered independently from the global film thickening[4]. Based on the fraction of each element accumulating in the film, the atomic percent of each element can be compared to the expected oxide compositions based on phase separation versus solute capture.

The current density contribution, $i_M(t)$, for each specific dissolved cation species, $M^+$, can be calculated from on-line ICP measurement[4]:

$$C_M = k(I_\lambda - I_\lambda^o)$$

$$i_M(t) = \frac{nFfC_M}{M_M}$$

$$i_{ICP} = \sum i_M$$

where $I_\lambda$ is the measured emission intensity of a given element, M, $I_\lambda^o$ is the background intensity, $k$ is a proportionality constant, C is the concentration of element M in the electrolyte stream, n is the electrons transferred per mole of M, F is Faraday's constant in coulombs/equivalent, $f$ is the electrolyte flow rate, $M_M$ is the molar mass of element M, and $i_{ICP}$ is given as the sum of all current densities associated with all dissolved species. The net current density contribution for oxide growth, $i_{ox}$, and the related oxide thickness, $x$, as well as the accumulated area density of elements collected into the oxide, $j_M$, are given as:

$$i_{ox} = i_{EC} - i_{ICP}$$

$$x = \frac{M_M}{nF\rho} \int_0^t i_{ox}\, dt$$

$$j_M = \frac{N_A}{nF} \int_0^t i_{ox}^M\, dt$$

where $\rho$ is the density of the oxide layer, $N_A$ is Avogadro's number, and $i_{ox}^M$ is the element-specific oxidation contribution for M related to congruent alloy dissolution and given as[4]:

$$i_{ox}^M = (at\%\, M)i_{EC} - i_{ICP}^M$$

A general schematic for the application of *operando* AESEC and DC measurements on electrochemical passivation reaction rates is given in Fig. S2. The results obtained for the accumulation of atoms in the film during passivation in the acidic NaCl solution are shown in Fig. S3 along with the overall film concentrations obtained for individual elements (Fig. S4 and Fig. S5) as derived from the atomic accumulation results. According to the Fig. S3, there is sufficient area density of Ni atoms at early times to form a continuous film layer. The critical value of approximately $10^{15}$ atoms/cm$^2$ is reached after 10 s for NiO formation, whereas the densities of Cr and Mo remain below the threshold, indicating that they are either captured in the Ni-rich film until Cr accumulation is sufficient around 1,000 s for the growth of another Cr-rich film or exist as phase-separated islands or oxide nanoparticles. From the calculated atomic concentrations during electrochemical passivation of the alloys shown in Figs S4 and S5, the film is primarily Ni-rich for all times, eventually becoming increasingly enriched with uncaptured Cr starting around 1,000 s. The expected concentration of Ni, Cr, and O for two possible phase-separated film compositions are shown in Figs. S3 and S4: 50% NiO + 50% Cr$_2$O$_3$ (Fig. S4) or NiO + NiCrO$_3$ with 25% of all Cr captured in the NiO which forms early and the remaining 75% of Cr atoms forming metastable NiCrO$_3$ (Fig. S5). The solute-captured condition (NiO + NiCrO$_3$) is much more consistent with film concentration data derived experimentally from the AESEC method, where at long times the concentrations approaches those expected for 50% NiO + trapped Cr and 50%

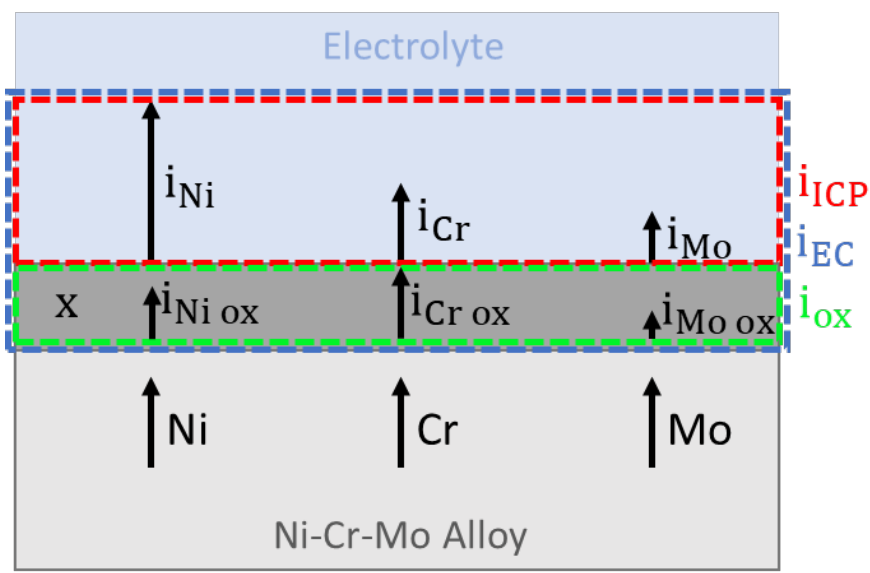

NiCrO$_3$.

**Fig. S2.** General schematic for the application of *operando* AESEC and DC measurements to electrochemical passivation reaction rates.

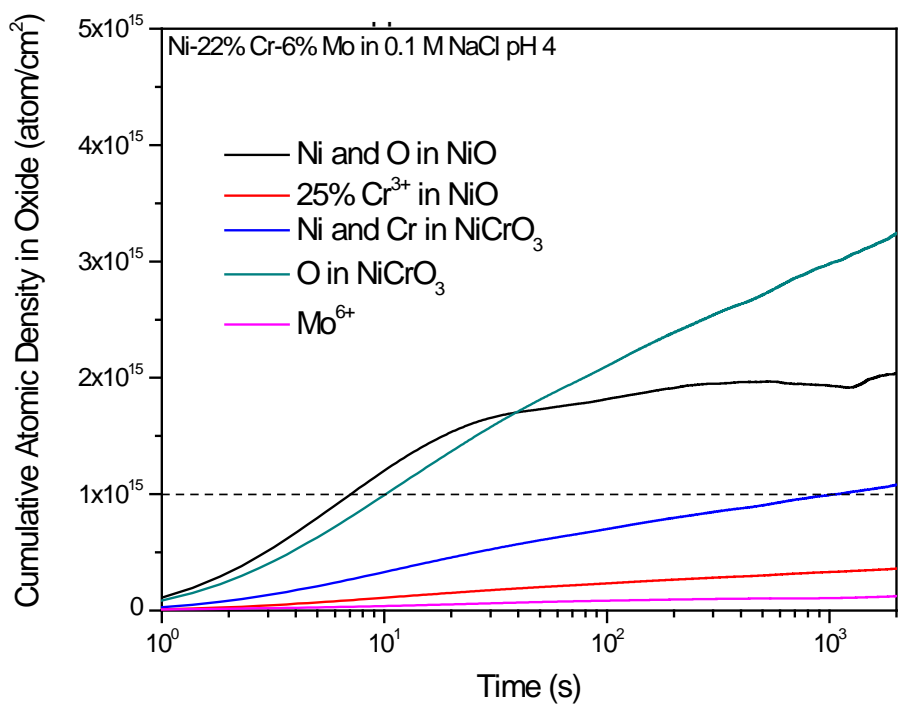

**Fig. S3.** The accumulation of atoms in the film during passivation in the acidic NaCl solution.

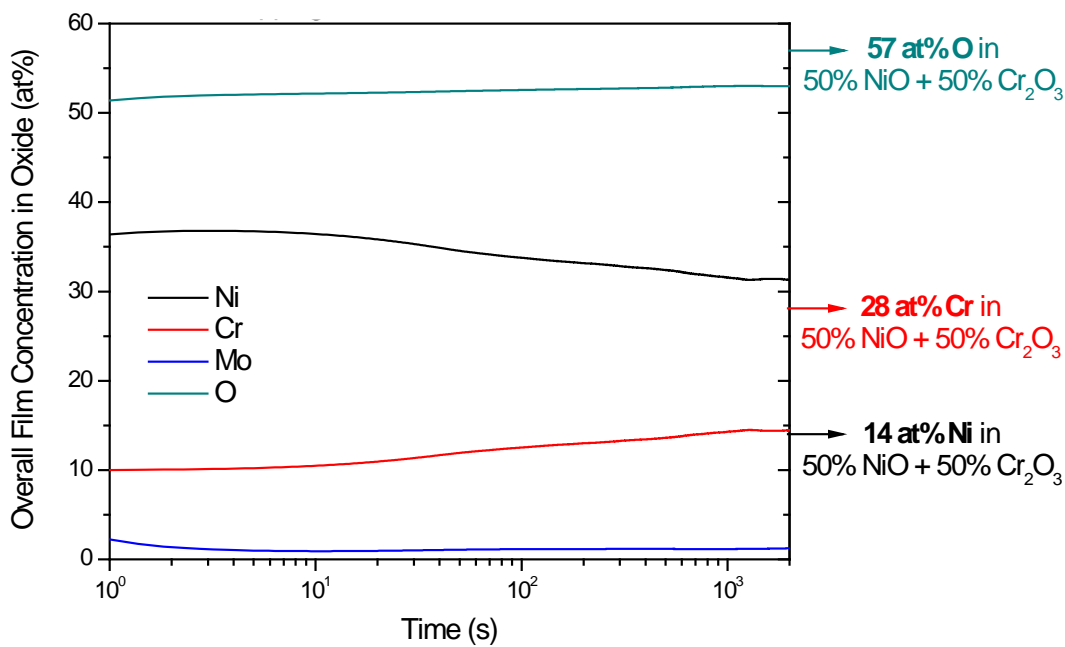

**Fig. S4.** Plot of the concentrations of the different species as color coded in the legend from the electrochemical data, assuming that only the NiO and $Cr_2O_3$ phases are formed. As shown on the right, the concentrations are very different from what one needs using the known composition of the metal alloy.

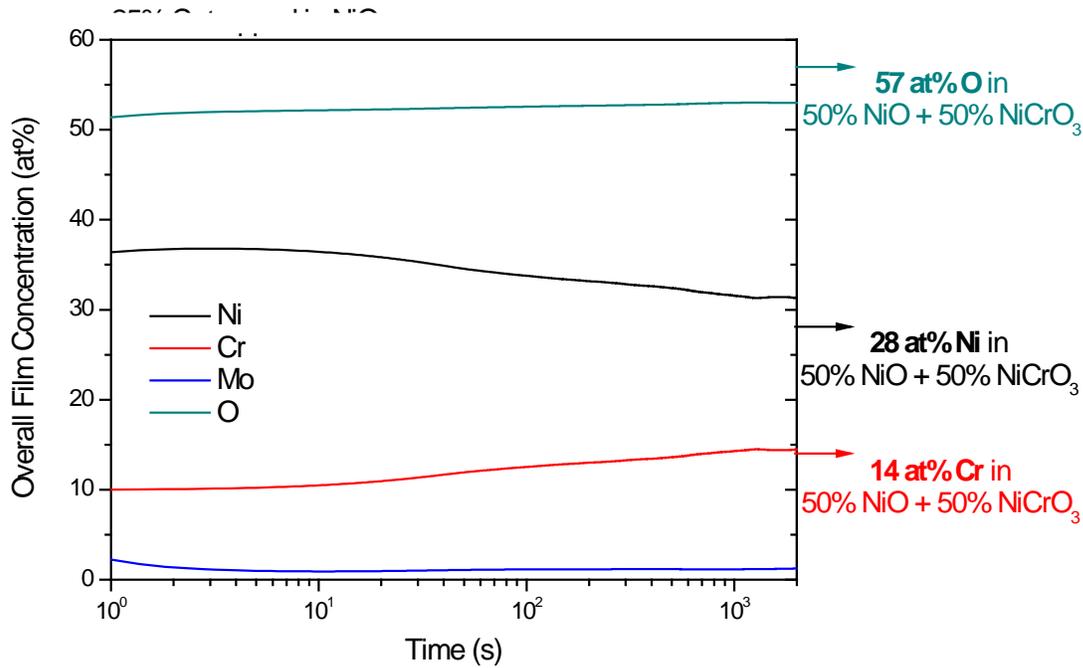

**Fig. S5.** Plot of the concentrations of the different species as color coded in the legend from the electrochemical data, assuming that the solute trapped phases determined experimentally are formed. As shown on the right, the concentrations are semi-quantitatively consistent with the known composition of the metal alloy.

## 3. Thermodynamic Condition for Solute Trapping and Non-equilibrium Solute Capture

We will first briefly describe conventional solute trapping. The moving interface between two phases during a phase transformation may be assumed to be at local equilibrium when the interfacial velocity is small compared to the atomic diffusivities in the material. For large relative interfacial velocities, however, this assumption of local equilibrium can break down. Interfacial equilibrium requires solute redistribution across the interface, which will not have time to occur if the interface is moving too quickly. During the rapid solidification of alloys, for example, a solid can form with a solute composition approaching that of the melt rather than its equilibrium composition[5,6]. In this case the solute is said to be "trapped" in the solid because its chemical potential increases in the solid relative to the melt. The increase in solute chemical potential is thermodynamically permitted so long as the total free energy change of the transformation per mole of material transferred from liquid to solid, $\Delta G = \Sigma_i [c_i (\mu_i^S - \mu_i^L)]$ where $c_i$ and $\mu_i$ are the mole fraction and chemical potential of species i, is negative. The general conditions for solute trapping are that the thermodynamic condition given above is satisfied and that the interface velocity v is large compared to the solute diffusion velocity. Strictly this is an effective interfacial diffusion constant (e.g.[7-9]) divided by some effective hopping distance. Current evidence indicates that the diffusion constant is likely bounded by the bulk solute diffusion coefficients of metal and oxide, and the hopping distance is often related to a width of an intrinsically rough interface (e.g.[10-14] and references therein). Note that for solute trapping, the interface velocity is in general proportional to the change in free energy of forming oxide of a given composition from metal of a different composition.

Non-equilibrium solute capture is different. The interface may be physically static, with a net flux of vacancies (nickel for nickel oxide) across it as well as (depending upon the oxide and metal) other species such as oxygen vacancies or metal interstitials. The velocity term in solute trapping is replaced by the sum of any physical interface motion and the net flux across the interface[15]. The relevant diffusion constant will be an effective interface term for a vacancy to diffuse along the interface (probably in the metal) and then jump back into the oxide to enable cation exchange, as against the vacancy diffusing into the semi-infinite sink of the metal. This will be comparable to a Zener-type cation redistribution in oxides (e.g. [16-21]), but now taking place across the interface with diffusion parameters which may be very different from those of the bulk materials. We will later approximate this using the oxide diffusion constant and a hopping distance of 1nm which are reasonable estimates. In addition to exchange there will also be a term which can be described as an effective velocity for nucleation of phase separated oxides; the velocity for equilibrate that competes with the growth velocity of the oxide will be the sum of both this phase separation velocity and one analogous to the term in solute capture.

Being more specific, to repeat what is described in the main text, the important measure of the extent of non-equilibrium solute capture in an effective medium model is $\beta = v^{Eff}/v^{Eq}$, where $v^{Eff}$ is the effective velocity of the interface combining both physical motion and the flux of atoms, and $v^{Eq}$ is an effective velocity for equilibration. The later will be the sum of a two terms, $v^{Eq} = v^{Exch} + v^{Nuc}$. The first, $v^{Exch}$, is similar to that in solute trapping $v^{Exch} = a/D_i$ where $D_i$ is an interface diffusion coefficient for a Zener exchange of the relevant atoms and a is the hopping distance for exchange of atoms between the oxide and metal to equilibrate the composition of the oxide. The second will be an effective velocity for nucleation of phase separation, for instance rocksalt $Ni_{1-x}Cr_{2x/3}O$ phase separating into rocksalt NiO and corundum $Cr_2O_3$ doped within solubility limits.

To evaluate whether capture is thermodynamically possible, we computed $\Delta G = \Sigma_i [c_i^M (\mu_i^{Ox} - \mu_i^M)]$, where $c_i$ is the composition of the metal alloy, and $\mu_i$ are the mole fraction and chemical potential of the species in the oxide and metal, respectively. If $\Delta G < 0$ then it is thermodynamically possible to form an oxide of composition $c_i^{Ox}$ from a metal of compositions, $c_i^M$. We consider the oxidation of a $NiCr_{0.05}$ alloy at 700 C and atmospheric pressure as a test case. The thermodynamics were evaluated by plotting the free energies of the alloy and rocksalt, corundum, and spinel oxide crystallographic phases per mole of Ni and Cr as a function of the Cr mole fraction in Fig. S6.

These free energy curves were calculated from the thermodynamic assessment of Taylor and Dinsdale[22]. Chromium defects in the rocksalt were considered as $Cr^{3+}$ and nickel defects in the corundum treated as $Ni^{2+}$. We reduced the free energy to a function of only the Cr mole fraction on the cation sublattice with the assumptions of structure conservation, charge neutrality, and that the oxygen in the oxide is in equilibrium with oxygen gas.

As a cross-validation, DFT calculations were performed with the all-electron augmented plane wave + local orbitals WIEN2K code[23] for $Ni_{1-x}Cr_xO_{1+x/2}$ compositions with $Ni^{2+}$ and $Cr^{3+}$ in either rocksalt or corundum structures as also shown in Figure S6 as well as the known spinel $NiCr_2O_4$. In these calculations it was found that the substitutions were lowest in energy when they followed the antiferromagnetic ordering of the parent rocksalt NiO or corundum $Cr_2O_3$. For the spinel, a search over antiferromagnetic orderings was performed, as this is not well defined in the current literature. The lowest energy configuration was ordered on the twin-related (100) or (010) planes for both $Ni^{2+}$ or $Ni^{3+}$ and $Cr^{3+}$. Technical parameters for these calculations were use of the PBE functional[24] exchange-correlation potential, the SCAN[25] exchange-correlation energies, a hybrid fraction of 0.25 for the d-electrons of both Ni and Cr using an on-site approach[26-28], oversampling of the exchange-correlation and otherwise somewhat conventional parameters. To avoid inconsistencies with the treatment of the metal reference in the DFT calculations, the known heats of formation for nickel oxide and chromia were used as references, for Figure S6 the values at temperature are used. This assumes that the relevant entropies can be approximated by a rule of mixtures. Additional calculations with $Ni^{3+}$ and $Cr^{2+}$ were performed, but these were significantly higher in energy and are not included here; full details will be described elsewhere.

At thermodynamic equilibrium there are only three possible phases: nickel oxide NiO, chromia $Cr_2O_3$ and possibly the spinel $NiCrO_3$. There is too much scatter in the available thermodynamic data for the spinel[29] for conclusive statements to be made. Experimentally we rarely saw it, suggesting that it is either metastable or marginally stable. In addition to these stable phases there are (insulating) metastable solid solutions of $Ni_{1-x}Cr_xO_{1+x/2}$ based upon both rocksalt and corundum structures across the whole composition range of x. We note that the calculations are consistent with experimental work attempting to produce a half-metal distorted perovskite NiCrO[30,31]. (Our DFT calculations indicated that the structure with antiferromagnetic ordering was an insulator with lower energy than the hypothetical half-metal considered other work[32,33].)

Chromia $Cr_2O_3$, has a larger thermodynamic driving force for formation than the rocksalt oxide NiO, but the rocksalt often forms initially in experiments. This is likely due to the smaller degree of structural reconfiguration required to form rocksalt from the FCC lattice of the alloy; the rocksalt structure can be formed directly from FCC by occupying the octahedral interstices with oxygen atoms, and allowing the lattice to expand. In the limit of very rapid oxidation, we would expect the rocksalt to form at the alloy composition. With decreasing oxidation rate, the chromium content in the oxide should increase because chromium has a much larger oxidation potential than nickel. From Fig. S6, we see that that the tangent line drawn to the free energy curve of the metal for all compositions of the alloy lies above the free energy of the oxide, indicating that ΔG < 0 for any $c_{cr}^M$. In fact, it is thermodynamically possible to form the rocksalt of *any* chromium content, all the way up to pure CrO, from the NiCr$_{0.05}$ alloy.

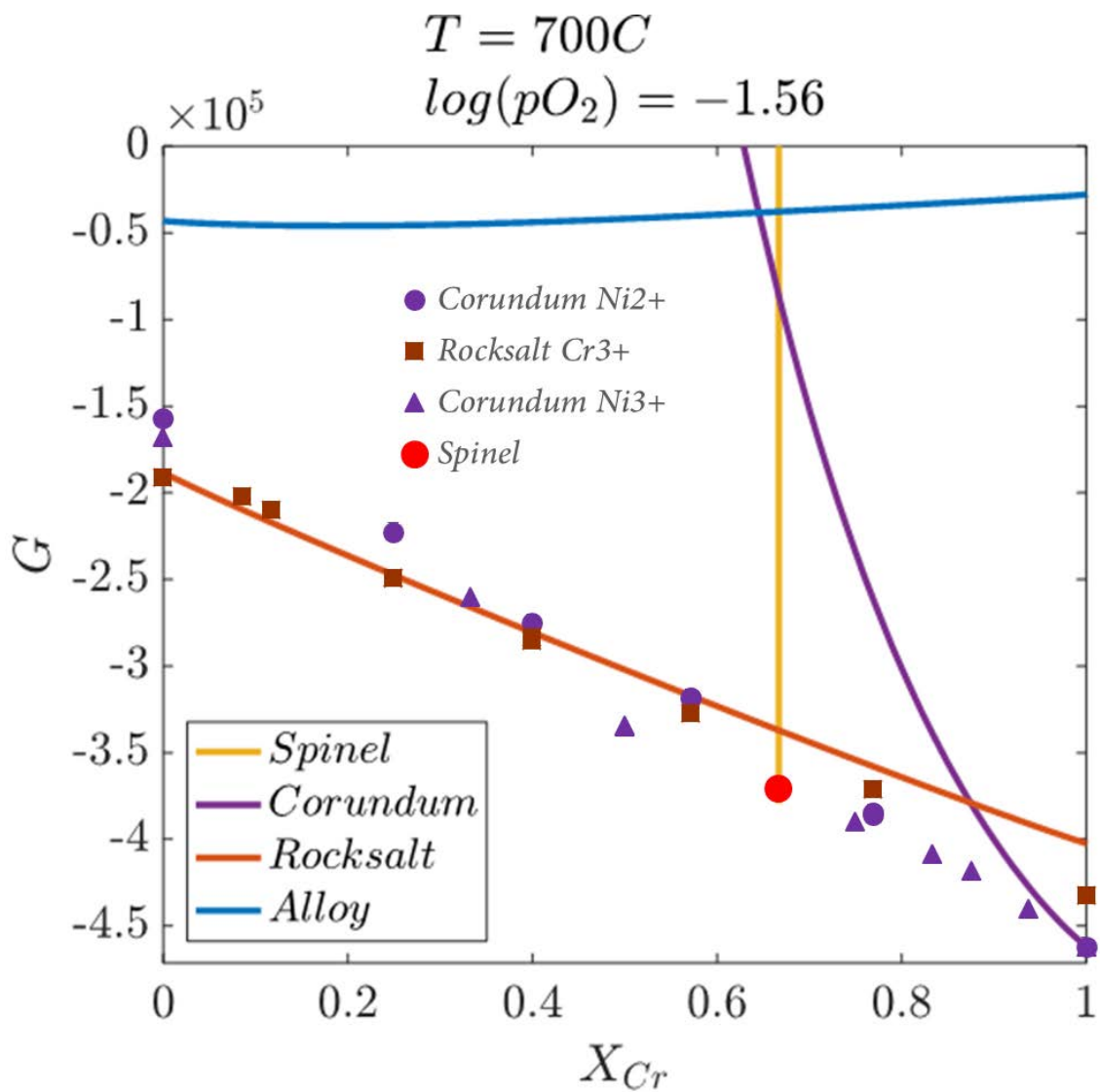

**Fig. S6.** Plot of the free energy of the different crystallographic phases in J/mol as color coded, as a function of Cr composition assuming that the valences are always Ni$^{2+}$ and Cr$^{3+}$ with the oxygen content adjusted to achieve valence neutrality as appropriate. Points are from the DFT calculations, lines from the thermodynamic database.

**4 Kinetic Condition for Solute Trapping and Non-Equilibrium Solute Capture**

In the general case, the partitioning across a moving interface is well documented for the solidification case (e.g.[5,7,34]) and has been evaluated for a at least one related cases, namely during thin film growth[35]. The partitioning of a solute between the liquid and solid as a function of the velocity of the interface ($v$) can be written as[7]

$$k(v) = \frac{k_0 + \beta}{1 + \beta}$$

where $k_0$ is the equilibrium partitioning and $\beta = va/D$ where a is a diffusion hop distance which is typically taken as comparable to an interatomic spacing, and D is the diffusion constant as a function of temperature and other conditions which for solute trapping is normally that of the liquid phase. If $va/D \gg 1$, solute atoms are trapped during solidification with a composition far from $k_0$.

We argue that the same holds approximately for non-equilibrium solute capture, with appropriate changes of the relevant terms since the atomic processes are different. For instance, for a stationary interface, unlike that discussed above, fluxes through the interface can generate a jump in the chemical potentials, and hence non-equilibrium interfacial compositions. Using the results of [15] in a binary alloy, and neglecting off-diagonal terms in the coupling matrix, we find that the jump in the difference of chemical potential across the interface is, $\Delta(\mu_1 - \mu_2) = -bj$, where $b$ is a positive constant related to the interfacial diffusion coefficient, and $j$ is the flux of a component through the interface. At equilibrium, $j = 0$ and the usual equality of chemical potentials results. If, however, there is a flux through the stationary interface, then the jump is nonzero, and thus the compositions at the interfaces are not those given by the usual equilibrium phase diagram.

Because of the differences, we need to redefine the relevant terms. The same overall formulation holds, with the slightly different definition of $\beta = v^{Eff}/v^{Eq}$, where $v^{Eff}$ is the effective velocity of the interface combining both physical motion and the flux of atoms, and $v^{Eq}$ is an effective velocity for equilibration as discussed above.

To demonstrate feasibility, we will use for oxidation conventional values from the literature and ignore the nucleation velocity. The hardest number to estimate is the interfacial diffusion constant $D_i$. Since exchange of (metal) atoms involves diffusion into the oxide, we used a more conservative estimate based upon vacancy diffusion rates in the oxide for the high temperature case. For instance, at 800 C the diffusivity of chromium in NiO is $2 \times 10^{-20}$ cm$^2$/s[36]. If we take $a$ as 1 nm, then the value of $a/D_i$ is $(2 \times 10^{-11}$ m/s$)^{-1}$. This implies that if the oxide growth rate significantly exceeds $2 \times 10^{-11}$ m/s then we would expect the rocksalt composition to approach the alloy composition. A more comprehensive plot is shown in Fig. 5A.

For electrochemical treatments, the effective velocity of the interface can be directly measured by the current, with a linear relationship from Faraday's law. Again neglecting the nucleation velocity, if we use the diffusion constant for the oxide extrapolated to room temperature we find that solute capture always occurs; we therefore used a larger value that includes a field-assisted term. (Physically this may correlate to role of the nucleation velocity within an effective medium approach, as this will probably be faster at the lower temperatures of electrochemical corrosion.) Being specific, we used a hopping distance of 1nm and assumed a field affected diffusivity of $10^{-20}$ cm$^2$/sec. The evolution of the protective oxide films in aqueous conditions is not a simple homogeneous process, rather a plethora of different processes taking place sometimes simultaneously, sometimes at different times or at different locations in the samples. These range from quasi-equilibrium with almost all effective interface velocities up to much faster processes where there is a breakdown of the film locally leading to pitting. The results are shown in Fig. 5B.

## 5. Density of States of Doped Oxides

To probe the effect of solute capture, two simple cases were considered: Cr in NiO and Mo in NiO, density functional theory calculations were performed with the all-electron augmented plane wave + local orbitals WIEN2K code[23] using the PBEsol[37] functional with on-site hybrid corrections[26-28] of 0.25 for the metal atoms. Lattice parameters were those for the DFT relaxed bulk structures. For the calculations muffin-tin radii of 1.9, 1.7, 1.9 and 1.9 au were used for the Ni, O, Cr and Mo atoms respectively, the plane wave extension (RKMAX) values was 6.5, and a Brillouin-Zone sampling of 8x8x8.

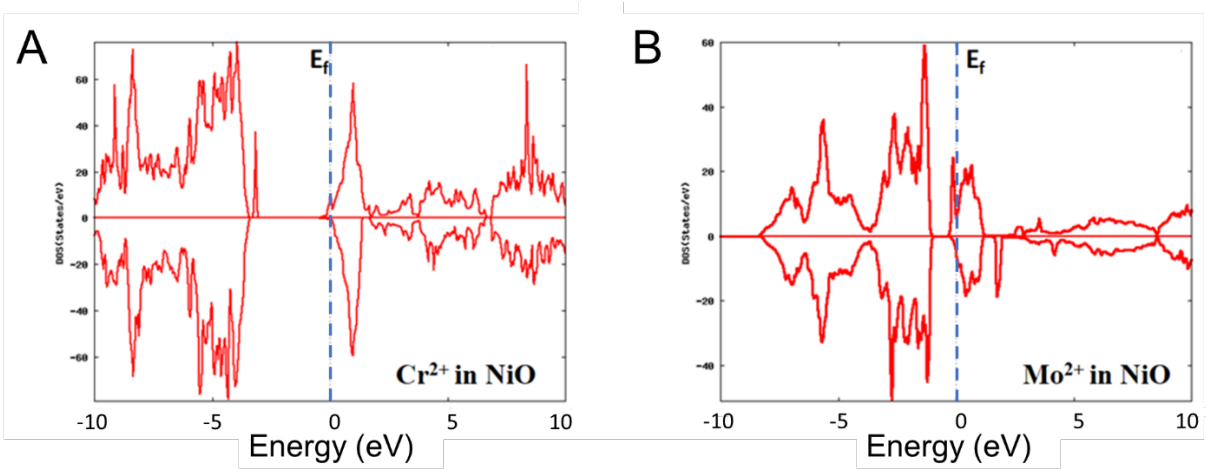

Fig. S7. Total density of states of (A) Cr and (B) Mo doping in a 2×2×2 rocksalt NiO supercell with the charge state 2$^+$, calculated by WIEN2K using an on-site hybrid functional. The dopant atoms introduce states in the gap of the base oxide, which are going to lead to doped semiconductor behavior of the oxide film, which will change how potentials develop across the oxide film.

**Supplemental References**

1       Ogle, K. & Weber, S., Anodic Dissolution of 304 Stainless Steel Using Atomic Emission Spectroelectrochemistry. *Journal of The Electrochemical Society* 147 (5), 1770 (2000).
2       Hamm, D., Ogle, K., Olsson, C.O.A., Weber, S., & Landolt, D., Passivation of Fe–Cr alloys studied with ICP-AES and EQCM. *Corrosion Science* 44 (7), 1443-1456 (2002).

3. Duarte, M.J. *et al.*, Element-resolved corrosion analysis of stainless-type glass-forming steels. *Science* 341 (6144), 372-376 (2013).
4. Lutton, K., Gusieva, K., Ott, N., Birbilis, N., & Scully, J.R., Understanding multi-element alloy passivation in acidic solutions using operando methods. *Electrochem Commun* 80, 44-47 (2017).
5. Baker, J.C. & Cahn, J.W., Solute Trapping by Rapid Solidification. *Acta Metall Mater* 17 (5), 575-578 (1969).
6. Boettinger, W.J., Coriell, S.R., & Sekerka, R.F., Mechanisms of Microsegregation-Free Solidification. *Materials Science and Engineering* 65 (1), 27-36 (1984).
7. Aziz, M.J., Model for Solute Redistribution during Rapid Solidification. *Journal of Applied Physics* 53 (2), 1158-1168 (1982).
8. Aziz, M.J., Dissipation-Theory Treatment of the Transition from Diffusion-Controlled to Diffusionless Solidification. *Appl Phys Lett* 43 (6), 552-554 (1983).
9. Aziz, M.J. & Kaplan, T., Continuous Growth-Model for Interface Motion during Alloy Solidification. *Acta Metall Mater* 36 (8), 2335-2347 (1988).
10. Asta, M. *et al.*, Solidification microstructures and solid-state parallels: Recent developments, future directions. *Acta Materialia* 57 (4), 941-971 (2009).
11. Mishin, Y., Asta, M., & Li, J., Atomistic modeling of interfaces and their impact on microstructure and properties. *Acta Materialia* 58 (4), 1117-1151 (2010).
12. Yang, Y. *et al.*, Atomistic simulations of nonequilibrium crystal-growth kinetics from alloy melts. *Phys Rev Lett* 107 (2), 025505 (2011).
13. Kuhn, P. & Horbach, J., Molecular dynamics simulation of crystal growth in Al50Ni50: The generation of defects. *Physical Review B* 87 (1), 5 (2013).
14. Xu, C. *et al.*, Atomistic simulations of solidification process in B2-LiPb solid(001)-liquid system. *J Cryst Growth* 470, 113-121 (2017).
15. Gurtin, M.E. & Voorhees, P.W., The thermodynamics of evolving interfaces far from equilibrium. *Acta Materialia* 44 (1), 235-247 (1996).
16. Walters, D.S. & Wirtz, G.P., Kinetics of Cation Ordering in Magnesium Ferrite. *Journal of the American Ceramic Society* 54 (11), 563-& (1971).
17. Becker, K.D. & Rau, F., High-Temperature Ligand-Field Spectra in Spinels - Cation Disorder and Cation Kinetics in Nial2o4. *Berichte Der Bunsen-Gesellschaft-Physical Chemistry Chemical Physics* 91 (11), 1279-1282 (1987).
18. Sujata, K. & Mason, T.O., Kinetics of Cation Redistribution in Ferrospinels. *Journal of the American Ceramic Society* 75 (3), 557-562 (1992).
19. Kashii, N., Maekawa, H., & Hinatsu, Y., Dynamics of the cation mixing of MgAl2O4 and ZnAl2O4 spinel. *Journal of the American Ceramic Society* 82 (7), 1844-1848 (1999).
20. Redfern, S.A.T., Harrison, R.J., O'Neill, H.S.C., & Wood, D.R.R., Thermodynamics and kinetics of cation ordering in MgAl2O4 spinel up to 1600 degrees C from in situ neutron diffraction. *American Mineralogist* 84 (3), 299-310 (1999).
21. Shi, J.M. & Becker, K.D., Kinetics and thermodynamics of cation site-exchange reaction in olivines. *Solid State Ionics* 181 (11-12), 473-478 (2010).
22. Taylor, J.R. & Dinsdale, A.T., A Thermodynamic Assessment of the Ni-O, Cr-O and Cr-Ni-O Systems Using the Ionic Liquid and Compound Energy Models. *Z. Metallk.* 81 (5), 354-366 (1990).
23. Blaha, P., Schwarz, K., Madsen, G., Kvasnicka, D., & Luitz, J., Wien2k, an augmented plane wave + local orbitals program for calculating crystal properties (Techn. Universität Wien, Austria, 2001).
24. Perdew, J.P., Burke, K., & Ernzerhof, M., Generalized Gradient Approximation Made Simple. *Phys Rev Lett* 77 (18), 3865-3868 (1996).

25  Sun, J., Ruzsinszky, A., & Perdew, J.P., Strongly Constrained and Appropriately Normed Semilocal Density Functional. *Phys Rev Lett* 115 (3), 036402 (2015).

26  Novak, P., Kunes, J., Chaput, L., & Pickett, W.E., Exact exchange for correlated electrons. *Phys Status Solidi B* 243 (3), 563-572 (2006).

27  Tran, F., Blaha, P., Schwarz, K., & Novak, P., Hybrid exchange-correlation energy functionals for strongly correlated electrons: Applications to transition-metal monoxides. *Physical Review B* 74 (15), 155108 (2006).

28  Tran, F. *et al.*, Force calculation for orbital-dependent potentials with FP-(L)APW plus lo basis sets. *Computer Physics Communications* 179 (11), 784-790 (2008).

29  Kurepin, V.A., Kulik, D.A., Hiltpold, A., & Nicolet, M., Report No. 02-04, Paul Scherrer Institute, Villigen, Switzerland, 2002.

30  Kusmartseva, A.F., Arevalo-Lopez, A.M., Halder, M., & Attfield, J.P., Bistability and relaxor ferrimagnetism in off-stoichiometric NiCrO3. *J Magn Magn Mater* 443, 293-299 (2017).

31  Chamberland, B.L. & Cloud, W.H., Preparation and Properties of NiCrO3. *Journal of Applied Physics* 40 (1), 434-435 (1969).

32  Qian, Y. *et al.*, Tuning the physical properties of antiferromagnetic perovskite oxide NiCrO3 by high-pressure from density-functional calculations. *Solid State Communications* 170, 24-29 (2013).

33  Lee, K.W. & Pickett, W.E., Compensated half-metallicity in the trigonally distorted perovskite NiCrO3. *Physical Review B* 83 (18), 180406 (2011).

34  Baker, J.C. & Cahn, J.W., Thermodynamics of Solidification in *ASM Semin. series: Solidification* (American Society for Metals, Metals Park, OH, 1971), pp. 23-58.

35  Han, X. & Spencer, B.J., A nonlinear model for surface segregation and solute trapping during planar film growth. *Journal of Applied Physics* 101 (8), 084302 (2007).

36  Chen, W.K., Peterson, N.L., & Robinson, L.C., Chromium Tracer Diffusion in NiO Crystals. *Journal of Physics and Chemistry of Solids* 34 (4), 705-709 (1973).

37  Perdew, J.P. *et al.*, Restoring the density-gradient expansion for exchange in solids and surfaces. *Phys Rev Lett* 100 (13), 136406 (2008).